\documentclass{article}
\usepackage{spconf,amsmath,graphicx,multirow}

\title{Study of positional encoding approaches for \\audio spectrogram transformers}
\name{Leonardo Pepino $^{\star\dagger}\qquad$ Pablo Riera $^{\star}\qquad$ Luciana Ferrer $^{\star}$ \vspace{-0.3cm} \thanks{This work was supported by a Google Faculty Research Award, 2019, and an Amazon Research Award, 2019.
Correspondence: lpepino@dc.uba.ar}}
\address{
$^{\star}$Instituto de Investigaci\'on en Ciencias de la Computaci\'on (ICC), CONICET-UBA, Argentina \\
$^{\dagger}$Departamento de Computaci\'on, FCEyN, Universidad de Buenos Aires (UBA), Argentina}

\begin{document}
%
\maketitle
\begin{abstract}
Transformers have revolutionized the world of deep learning, specially in the field of natural language processing. Recently, the Audio Spectrogram Transformer (AST) was proposed for audio classification, leading to state of the art results in several datasets.
However,  
in order for ASTs to outperform  CNNs, pretraining with ImageNet is needed.
In this paper, we study one component of the AST, 
the positional encoding, and propose several variants 
to improve the performance of ASTs trained from scratch, without ImageNet pretraining. Our best model, which incorporates conditional positional encodings, significantly improves performance on Audioset and ESC-50 compared to the original AST. 
\end{abstract}
\vspace{-0.1cm}
\vspace{-0.0cm}
\section{Introduction}
\label{sec:intro}
\vspace{-0.2cm}

In the last few years, models based on attention mechanisms have gained traction in the field of deep learning, leading to impressive results in many fields like natural language processing \cite{devlin2018bert, brown2020language} and computer vision \cite{dosovitskiy2020image, ramesh2021zero}. The most successful of such models is the transformer, proposed in \cite{vaswani2017attention}.
Recently, transformers and other attention-based approaches have been incorporated into audio models for representation learning \cite{baevski2020wav2vec, verma2020framework} and automatic speech recognition \cite{gulati2020conformer, moritz2020streaming}. Also, convolution-free models purely based on transformers have been recently proposed for audio processing \cite{gong21b_interspeech,verma2021audio}.

Unlike convolutional neural networks (CNNs), transformers do not have a limited receptive field and can see the whole input at each layer when computing every output, allowing it to capture long-range dependencies. Moreover, contrary to recurrent neural networks, the length of the paths the
signals traverse in the network to learn these long-range dependencies is constant, 
making it more suitable for learning relationships that span large temporal contexts.
The basic transformer, though, has a disadvantage: the attention mechanism is permutation-invariant, making it suboptimal for sequential data as ordering information is not taken into account by the model. To overcome this problem, several strategies have been proposed to add information about the position of an element in a sequence. One of the most common positional encoding (PE) approaches is to add embeddings encoding the position to the input sequence. These positional embeddings
can be learned by the model \cite{devlin2018bert, chu2021conditional} or designed by hand \cite{vaswani2017attention,su2021roformer}.
Another
PE strategy is to use relative attention \cite{shaw2018self, huang2018music, press2021train}, where the distance between the query and the key is used in the computation of the attention weights.

One of the transformer-based models which has been successful for audio tasks is the Audio Spectrogram Transformer (AST) \cite{gong21b_interspeech}, a Vision Transformer (ViT) \cite{dosovitskiy2020image} trained with audio spectrograms instead of images. 
However, ViT, and transformers in general, require a large amount of data to outperform CNNs in computer vision tasks. 
This is likely because CNNs have strong inductive biases due to their locality and weight sharing, making them more suitable when modest amounts of training data are used. On the other hand, when there is enough data, using CNNs can be suboptimal as long-range dependencies cannot be easily captured by these models \cite{d2021convit}. As an example, authors in \cite{dosovitskiy2020image} found that when 9M images were used for training, a big CNN outperformed ViT, but when 90M+ images were used, ViT outperformed CNNs. These results highlight the importance of inductive biases to achieve good generalization
\cite{mitchell1980need, baxter2000model}, specially when data is limited. 
In the case of AST, the authors \cite{gong21b_interspeech} found it was crucial to initialize their model using a ViT model pretrained on ImageNet, despite the fact that they train on Audioset \cite{audioset}, which has around 2 million audios.
They also found that it was  important to initialize the positional embeddings with the ones from the pretrained model,
taking advantage of the 2D spatial knowledge learned from images.

In this work, we aim to improve the PE used in the AST for acoustic event detection, so that ImageNet pretraining is not as essential. Based on an analysis of the patterns learned by the original absolute PE and taking into account the structure of audio spectrograms, we modified and extended different relative and conditional PE (CPE) strategies proposed in the literature. Our best model, which uses CPE, outperforms the absolute PE used in AST both in Audioset and ESC-50 \cite{piczak2015dataset} datasets, and is on par with state of the art results in ESC-50 without the need to pretrain the AST in ImageNet. Source code is available at \url{https://github.com/habla-liaa/ast-pe}.

\section{Methods}
\vspace{-0.2cm}
In this work we propose to modify the absolute PE method used in the original AST, replacing it with different PE strategies which take into account the time-frequency structure of spectrograms, introducing helpful biases in the model. In the following subsections we describe the original AST approach and the different proposed PE approaches.

\vspace{-0.2cm}
\subsection{Audio spectrogram transformers (AST)}

\begin{figure}[tb]
\centering
\centerline{\includegraphics[width=7.5cm]{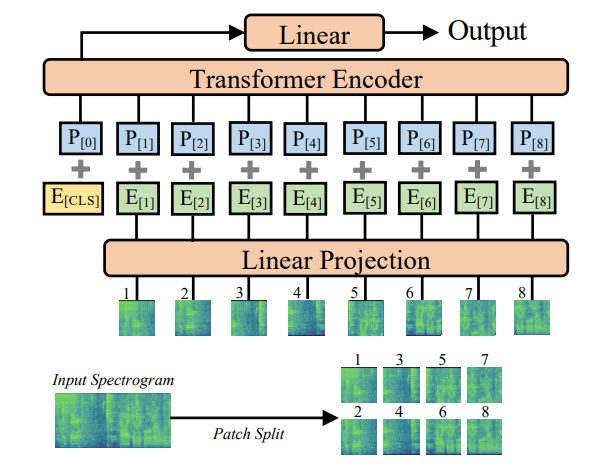}}
\vspace{-0.4cm}
\caption{AST architecture overview (taken from \cite{gong21b_interspeech}). The input spectrogram is rearranged into a sequence of patches and linearly projected to get embeddings $E_{[i]}$. Positional embeddings $P_{[i]}$ are summed to the sequence and fed to a transformer encoder.}
\label{fig:ast}
\end{figure}

\vspace{-0.2cm}

Audio spectrogram transformers (AST) \cite{gong21b_interspeech} have been recently proposed as a fully attentional model for audio classification. The architecture is inspired by visual transformers (ViT) \cite{dosovitskiy2020image}, but instead of taking images as input, it takes logarithmic melspectrograms 
extracted from audio signals.
The main idea behind these approaches, is that the image or spectrogram is split in rectangular patches, which are then concatenated into a sequence of patches and fed as input to a regular transformer, as shown in Figure \ref{fig:ast}.
Trainable absolute positional embeddings are added to the input patches to introduce position information in the model.
A [CLS] token is appended to the sequence and used to perform classification in a way similar to BERT \cite{devlin2018bert}. 

In the AST paper, the authors used patches of 16x16 and achieved state-of-the-art results in audio event detection, both in Audioset and ESC-50 \cite{piczak2015dataset} datasets, and in the Google Speech Commands dataset \cite{speechcommandsv2}. 
They showed that a crucial aspect to achieve these results was to pretrain the model using ImageNet. Training in Audioset from scratch, in spite of being a large-scale dataset with more than 2 million 10 seconds long audios, led to worse results than using the pretrained model for initialization. 
Moreover, they showed the importance of initializing the positional embeddings with the weights learned during the ImageNet pretraining. These results indicate that some of the patterns 
that were learned from images, are useful when working with audio spectrograms. We hypothesize that locality and translation invariance (specially in the time axis) might be important 
to help the model achieve better generalization.    

\subsection{Conditional positional encodings (CPE)}
\vspace{-0.1cm}

\begin{figure}[t]
\centering
\centerline{\includegraphics[width=\linewidth]{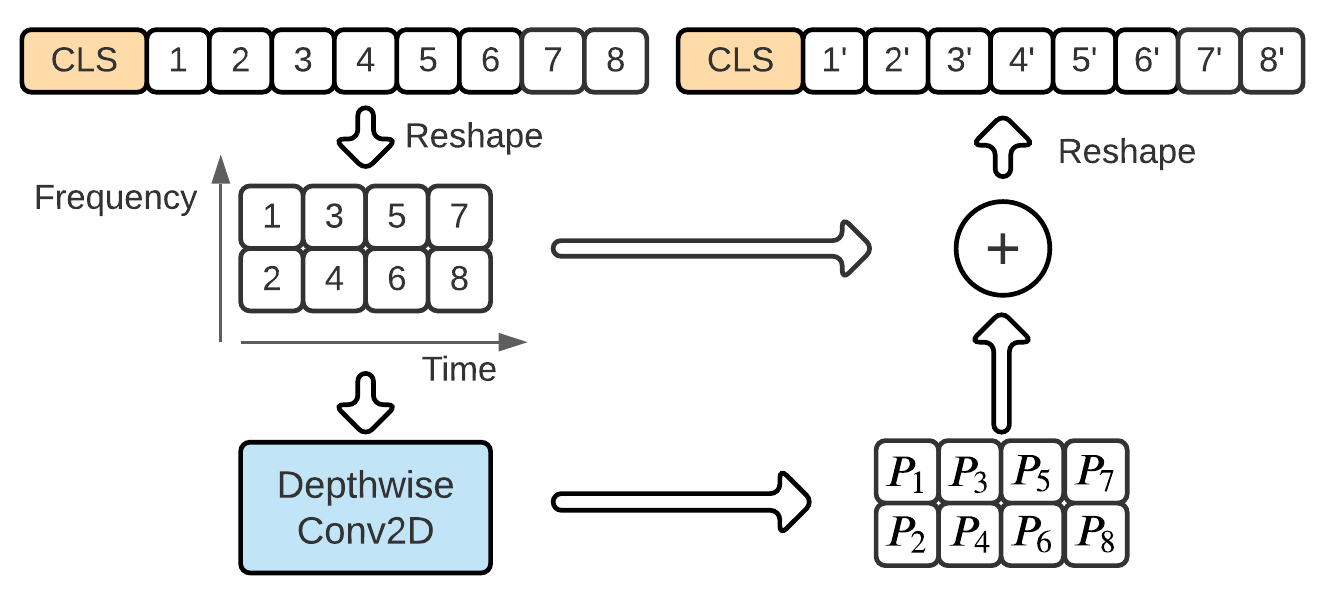}}
\vspace{-0.2cm}
\caption{Positional encoding generator used in CPVT. The sequence of patches is rearranged into a matrix $E$ with the original spectrogram shape and a depth-wise  convolution layer with kernel size 3x3 generates a new matrix of positional embeddings $P_i$. Then, as in the absolute PE, the $P$ and $E$ matrices are summed and the result is again rearranged into a sequence.}
\label{fig:peg}
\end{figure}

Conditional positional encoding for visual transformers (CPVT) \cite{chu2021conditional} has been recently proposed to favor translation invariance in ViT, improving the performance of the original model. Instead of learning a fixed set of positional embeddings, in CPVT these are dynamically generated and depend on the input sequence. By using a 2D convolutional layer as the positional encoding generator (PEG), the CPVT can keep translation invariance and adapt to arbitrary input sizes. The PEG block is shown in Figure \ref{fig:peg}. In \cite{chu2021conditional}, authors showed that placing the PEG layers at the output of the first 5 transformer blocks led to the best results. CPVT is very efficient, introducing only 38.4K extra trainable parameters.

\subsection{Relative attention}
\vspace{-0.1cm}

Relative attention models introduce information about the relative position between different components in a sequence by affecting the attention products with information about the distance between the queries and keys. 
In this work, we modified the relative attention mechanism proposed in \cite{huang2018music}, which learns embeddings for each possible distance between queries and keys, and is originally used to generate music in an autoregressive way. We removed the autoregressive constraint and 
incorporated relative positions not only in the time but also the frequency axis. The resulting scaled dot-product attention head is given by
\begin{equation}
\label{eq:huangatt}
    \mathrm{Att}(Q,K,V) = \mathrm{Softmax}(\frac{QK^T + R}{\sqrt{d_k}})V
\end{equation}
where Q, K and V are the query, keys and values matrices; $d_k$ is the size of the keys; and $R$ is the relative attention matrix with elements $R_{ij} = Q_{i}E^{t}_{\Delta t(i,j)} + Q_{i}E^{f}_{\Delta f(i,j)}$, where $E^{t}$ and $E^{f}$ are embeddings learned by the model for each possible distance between elements of the sequence in the time and frequency axes, respectively, and $\Delta t(i,j)$ and $\Delta f(i,j)$ return the difference in time and frequency between elements $i$ and $j$ of the sequence. In our implementation, the attention heads share the $E^{t}$ and $E^{f}$ matrices, but they are different for each transformer block. The main difference with the attention proposed in \cite{huang2018music} is that we added the term $Q_{i}E^{f}_{\Delta f(i,j)}$ to incorporate relative distances in the frequency axis. Note that this strategy is more efficient in terms of number of parameters than using absolute PE, adding 58.4k parameters compared to the 190.5k parameters of the absolute PE.

We also extended the Attention with Linear Biases (ALiBi) approach \cite{press2021train} to be non-autoregressive and use relative distances in both time and frequency. In ALiBi, the $R$ matrix is fixed, so it does not add any trainable parameter to the model, and encourages locality which might be desirable for modelling spectrograms. In our implementation, half of the heads in each layer use $R_{ij} = -m|\Delta t(i,j)|$, and the other half use $R_{ij} = -m|\Delta f(i,j)|$, where $m = 0.5^{16h/N_h}$, $h$ is the index indicating the attention head number, and $N_h$ is the number of attention heads in
that layer.
In consequence, as the distance between a given query and key increases, a larger number is subtracted from the $QK^T$ product, resulting in 
larger attention weights for the values that are in the neighbourhood of the query.
Also, as $m$ is different for each head, the heads with a larger $m$ will be more focused in the neighbourhoods of the query while the ones with a smaller $m$ will be less affected by the distance between query and key.

\section{Experimental setup}

We performed our experiments on the Audioset and ESC-50 datasets for the task of acoustic event detection. In this section we describe these datasets and the experimental setup used to obtain results in each case.

\subsection{Audioset experiments}
Audioset is a dataset with over 2 million 10-second audio clips from YouTube videos, which contain multi-label annotations of the sound events present in the clips out of 527 possible event classes. We downloaded 18026 audios from the balanced subset, 1637982 from the unbalanced subset, and 16710 from the evaluation subset, which correspond to about 80\% of the original lists for each set due to the rest of the audios being unavailable from YouTube at the time of download.
We train our models on the concatenation of balanced and unbalanced subsets, and report the mean Average Precision (mAP) obtained on the evaluation set. The mAP is given by the area under the precision-recall curve for each event class averaged over all the classes, and is a commonly used metric for object and audio event detection \cite{everingham2010pascal}.

The audio clips were resampled to 16 kHz and a log-melspectrogram was calculated using a Hann window with a length of 25 ms, a hop size of 10 ms, and 64 mel-frequency bins. The resulting spectrogram was scaled to a range between 0 and 1, using statistics from the whole training dataset. Specaugment \cite{park2019specaugment} was applied with a rate of 0.5 to the linear spectrogram (before applying the mel filter banks), using a maximum of 2 masks in each axis, with a maximum length of 64 frequency bins and 100 frames.

The log-melspectrogram is split into chunks in time and bands in frequency, obtaining patches which are then ordered into a sequence (Figure \ref{fig:peg}). In our case, 31 chunks of 32 frames and 8 bands of 8 mel coefficients are used, resulting in a sequence of 248 patches. 
The transformer consists of 12 blocks with an embedding dimension of 768 and 12 attention heads, as these are hyperparameters commonly used in the literature. Finally, the [CLS] token is used as input to a dense layer with sigmoid activations which maps the audio representation to the 527 Audioset classes.

The models were trained for 290k training steps, which corresponds roughly to 12 epochs, with a batch size of 64, and model parameters were saved every 10k steps. As the model with learned relative attention required more memory, we reduced its batch size to 32 and trained it for 470k steps. We used Adam optimizer in all our models, with a linear warm-up of the learning rate during the first 30k steps, and an exponential decay, reaching a maximum learning rate of 5e-4. As in \cite{gong2021psla}, stochastic weight averaging (SWA) \cite{izmailov2018averaging} was performed, averaging the weights corresponding to the last 10 saved models.

\subsection{ESC-50 experiments}

ESC-50 \cite{piczak2015dataset} is a dataset for environmental sound classification, consisting of 2000 5-second recordings organized in 50 classes. The dataset is class-balanced and contains animal, natural soundscapes and water, human non-speech, interior/domestic, and exterior/urban sounds.

We finetuned the models trained in Audioset after SWA, replacing the output layer, and changing its activation from sigmoid to softmax as ESC-50 is not a multilabel dataset. During the first 10 epochs, we trained only the output layer using a learning rate of 0.001, and then we unfreezed the whole model and kept training it for 40 epochs, with an initial learning rate of 1e-4 decaying it by a factor of 0.85 at each epoch.
We evaluated our models using 5-fold cross-validation, using the official folds and reporting the accuracy.

\section{Results and Discussion}

\begin{table}[]
\centering
\label{table:results}
\small

\begin{tabular}{lcc}
PE Method           & Audioset & ESC-50 \\ \hline
None             & 0.286  & 81.2   \\
Absolute         & 0.313 & 87.5   \\
ALiBi 2D         & 0.307 & 86.3   \\
Time ALiBi       & 0.319 & 87.6   \\
Learned Relative & 0.329 & 87.8   \\
Conditional              & 0.343 & \textbf{91.4}   \\
Conditional + Absolute   & \textbf{0.344} & 90.0   \\
\hline
AST \cite{gong21b_interspeech} & \textbf{0.485} & \textbf{95.7} \\
WEANET \cite{kumar2020sequential} & 0.398 & 94.1 \\
EfficientNet \cite{kim2020urban} & - & 89.5
\end{tabular}
\caption{Comparison of the performance obtained in Audioset and ESC-50 with the different proposed PE approaches. The values correspond to mAP in Audioset and to accuracy in ESC-50. We also show the results from state of the art models for comparison.}
\end{table}

\begin{figure}[tb]
\centering
\centerline{\includegraphics[width=0.9\linewidth]{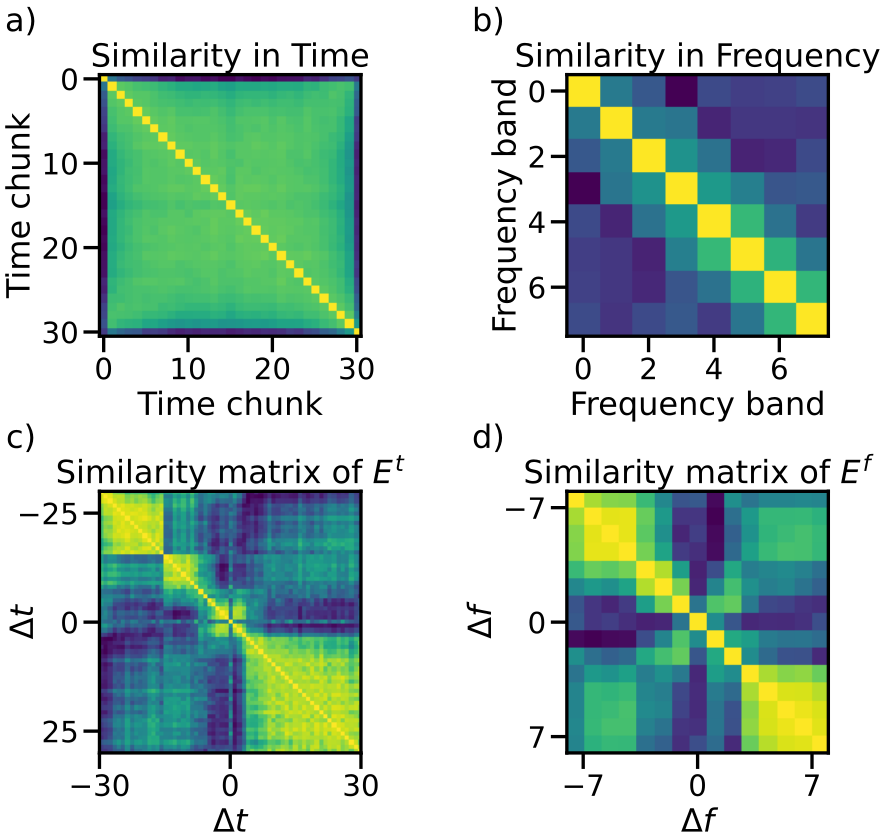}}
\vspace{-0.3cm}
\caption{Figures a) and b) show the cosine similarity matrices of the absolute positional embeddings. The similarity in time (frequency) is obtained by concatenating all the embeddings corresponding to the same time (frequency), and then calculating the pairwise cosine similarity. Figures c) and d) show the similarity of the vectors in $E^t$ and $E^f$ for the Learned Relative PE approach.}
\label{fig:abs_sim}
\end{figure}

The performance obtained in Audioset and ESC-50 for each of the PE under study can be seen in Table~\ref{table:results}. As expected, the worst results are obtained when positional information is ignored (None) although, on ESC-50, those results are similar to the performance achieved by humans \cite{piczak2015dataset} and by CNNs \cite{tokozume2018learning, zhu2018learning}.
Incorporating absolute PE improves the performance both in Audioset and ESC-50. 
This is the most common type of PE, and it is used in the original AST and ViT works. However, we found that the model, when using this PE method, tends to differentiate positions only in the frequency axis, as seen in Figure \ref{fig:abs_sim}. This suggests that for acoustic event detection, combining information from distant time-steps might not be essential for reaching reasonable performance, as many acoustic events are stationary or have a duration shorter than the patch size in time (320 ms). Yet, another hypothesis could be that absolute positional embeddings are not able to help the model effectively learn temporal relationships that could be useful for the task, unless the model is pretrained with very large amounts of data as in \cite{gong21b_interspeech}. As we will see next, in our results, non-absolute PE performs better, giving support to this hypothesis.

The proposed extension of ALiBi (ALiBi 2D), performed worse than absolute PE. We hypothesize that the locality bias in ALiBi might be important for the time axis, but detrimental if used in the frequency axis. 
Because of this, we experimented with absolute PE to discriminate frequency positions, 
and ALiBi to introduce only time-distance information in the attention heads. In this case, all the attention heads add $R_{ij} = -m|\Delta t(i,j)|$ when computing the softmax weights. 
The results indicate that this approach (Time ALiBi) improves over ALiBi 2D and Absolute PE. 

When the $R$ matrix is learned using our extension of the relative attention proposed in \cite{huang2018music} (Learned relative), the performance is further improved. We think that in this scenario where a decent amount of data is available for training, making the
$R$ matrix trainable gives an advantage over ALiBi which uses a fixed $R$ matrix. Figure \ref{fig:abs_sim}c shows that, in contrast to the absolute positional embeddings, the relative positional embeddings are able to differentiate time regions in the past, present and future. Moreover, in this model, the positional embeddings interact with the queries, which might give the model more capacity.

Finally, the best results are obtained when using CPE. In contrast to the other PE approaches under analysis, CPE is adaptive depending on the input signal itself, which might be an advantage, although it makes the generated positional embeddings harder to interpret. Despite not having explicit information about the absolute position, CPE outperforms absolute PE, and gives results close to the state of the art in ESC-50 dataset. Finally, we also tried 
summing absolute positional embeddings to the transformer input (CPE + Absolute) but no significant gains were observed. This suggests that absolute position information is not required or it can be learned by the CPE as shown in \cite{chu2021conditional}.

\section{Conclusions}

In this paper, we studied different approaches to incorporate positional information in Audio Spectrogram Transformers (AST). We showed that, with a careful design of the positional encoding (PE) component that takes into account the structure of audio spectrograms, performance can be boosted with respect to learning absolute positional embeddings from scratch. In particular, using conditional PE provides a 9.9\% and 4.5\% of relative improvement for Audioset and ESC-50, respectively. Yet, our results in Audioset are still worse than those obtained with ImageNet pretraining. We believe the remaining gap could be narrowed by adapting other components of the transformer to take into account the intrinsic patterns of audio signals.

\small
\bibliographystyle{IEEEbib}
\bibliography{strings,refs}

\end{document}